\documentclass[]{interact}

\usepackage{epstopdf}% To incorporate .eps illustrations using PDFLaTeX, etc.
\usepackage[caption=false]{subfig}% Support for small, `sub' figures and tables
\usepackage[natbibapa,nodoi]{apacite}% Citation support using apacite.sty. Commands using natbib.sty MUST be deactivated first!
\theoremstyle{plain}% Theorem-like structures provided by amsthm.sty

\theoremstyle{definition}

\theoremstyle{remark}

\begin{document}

\articletype{}% Specify the article type or omit as appropriate

\title{AI-Based Automated Speech Therapy Tools for persons with Speech
Sound Disorders: A Systematic Literature Review}

\author{
\name{Chinmoy Deka\textsuperscript{a}, Abhishek Shrivastava\textsuperscript{a}, Ajish K. Abraham\textsuperscript{b} \thanks{CONTACT Abhishek Shrivastava. Email: shri@iitg.ac.in}, Saurabh Nautiyal\textsuperscript{a} and Praveen Chauhan\textsuperscript{a}}
\affil{\textsuperscript{a}Indian Institute of Technology Guwahati, Guwahati, 781039, Assam, India; \textsuperscript{b}All India Institute of Speech and Hearing, Mysore, 570006, Karnataka, India}
}

\maketitle

\begin{abstract}
This paper presents a systematic literature review of published studies on AI-based
automated speech therapy tools for persons with speech sound disorders (SSD). The
COVID-19 pandemic has initiated the requirement for automated speech therapy
tools for persons with SSD making speech therapy accessible and affordable. However, there are no guidelines for designing such automated tools and their required
degree of automation compared to the conventional speech therapy given by Speech
Language Pathologists (SLPs). In this systematic review, we followed the PRISMA
framework to address four research questions: 1) what types of SSD do AI-based
automated speech therapy tools address, 2) what is the level of autonomy achieved
by such tools, 3) what are the different modes of intervention, and 4) how effective are such tools in comparison with the conventional mode of speech therapy.
An extensive search was conducted on digital libraries to find research papers relevant to our study from 2007 to 2022. The results show that AI-based automated
speech therapy tools for persons with SSD are increasingly gaining attention among
researchers. Articulation disorders were the most frequently addressed SSD based
on the reviewed papers. Further, our analysis shows that most researchers proposed
fully automated tools without considering the role of other stakeholders. Our review
indicates that mobile-based and gamified applications were the most frequent mode
of intervention. The results further show that only a few studies compared the effectiveness of such tools compared to the conventional mode of speech therapy. Our
paper presents the state-of-the-art in the field, contributes significant insights based
on the research questions, and provides suggestions for future research directions.
\end{abstract}


\section{Introduction}

\begin{figure}[htp]
    \centering
    \includegraphics[width=\textwidth]{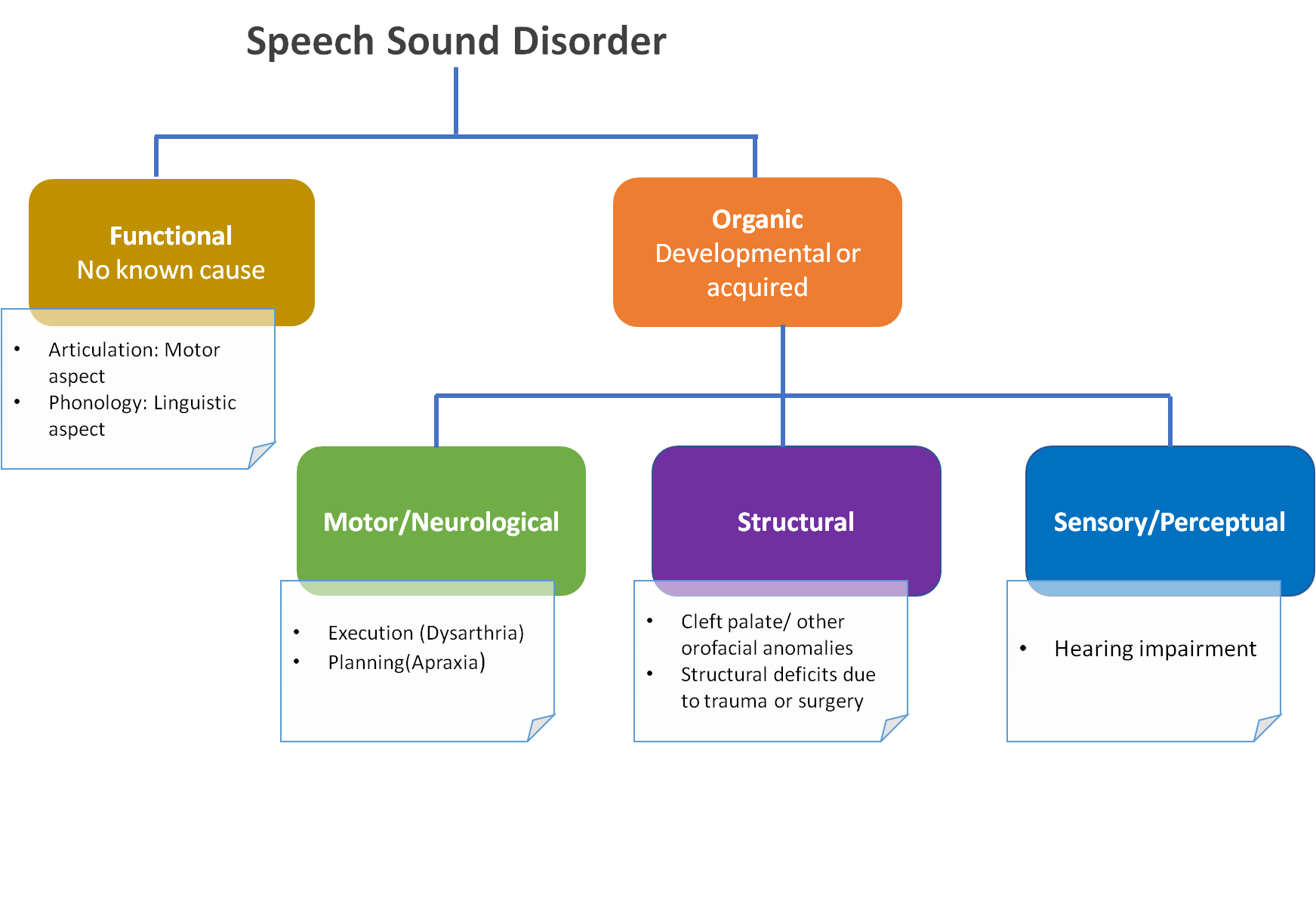}
    \caption{Classification of Speech Sound Disorders}
    \label{fig:classification_SSD}
\end{figure}

Speech Sound Disorder (SSD) refers to difficulties with perception, motor production,
phonological representation of speech sounds, and speech segments, which would cause
difficulties for the listener in perception (cite this). In short, a person with SSD finds it difficult to produce or use some sounds correctly. According to the American Speech Language
Hearing Association (ASHA), SSD can be organic and functional \citep{1_asha}. While organic
SSD result from an underlying motor/neurological, structural, or sensory/perceptual
cause, there is no known cause for functional SSD \citep{1_asha} (see Figure \ref{fig:classification_SSD}). The prevalence of SSD varies significantly according to different studies; however, these studies reflect the magnitude of the problem (cite this). Multiple studies have estimated that residual or persistent speech errors occur in 1\% to 2 \% of older children and adults \citep{2_flipsen2015}. In various studies overall, 2.3 \% to 24.6 \% of school aged children were estimated to have speech delay or speech sound disorders \citep{3_black2015communication,4_wren2016prevalence}. In a 2012 survey, National Center for Health Statistics found that 48.1 \% of 3 to 10 year old children and 24.44 \% of 11 to 17 year old children had speech sound problems \citep{3_black2015communication}. According to the 2011 census, in India, hearing impairment (18.9 \%) was the second leading disability, and speech impairment (7.5 \%) was the fifth highest disability \citep{5_velayutham2016prevalence}. In another survey conducted in India’s rural population, researchers found that around 6.07 \% were at risk of communication disorders, including speech sound disorder \citep{6_konadath2013prevalence}.
In addressing such speech impairments, Speech Language Pathologists (SLPs) play
a significant role in the screening, assessment, diagnosis, and treatment of persons with
SSD. Personalized speech therapy and practice monitored by SLPs can improve the
acquisition of speech skills \citep{7_duval2018spokeit}. However, the accessibility of SLPs is crucial for such intervention. A report suggests that up to 70 \% of SLPs have waiting lists, which indicates
a shortage in the workforce \citep{7_duval2018spokeit,8_robles2017onto}. Furthermore, according to United Nations Children’s Fund (UNICEF), there are not adequate speech language therapy services for children with communication disorders and disabilities \citep{9_lansdown2013children}. Moreover, speech therapy involves extended interactions and multiple sessions with SLPs. Such therapy requires extensive time, making it expensive and inaccessible for persons living in impoverished and rural areas. Addressing these issues of accessibility and expensiveness, many researchers have proposed AI-based automated tools for providing speech therapy autonomously to persons with SSD. With the advent of improved ASR tools, impaired speech datasets, and AI-based techniques, it is now feasible to build such autonomous tools for speech therapy. These autonomous tools embedded in mobile devices or provided as cloud services can be revolutionary in making speech therapy accessible and affordable. These tools will also aid in providing speech therapy through telepractice.

%  Add Figure 1

This paper reviews the research involving AI-based automated speech therapy tools
for persons with SSD during the last 15 years. In the next section, we present the
methodology adopted in this study. Section 3 reports the results and Section 4 presents
the discussion. Finally, we conclude the study in Section 5.

\section{Methodology}
We followed the PRISMA protocol to perform the systematic literature review to achieve higher transparency and reliability \citep{10_shamseer2015preferred}. We included studies that can cover one of the following questions we framed for this systematic literature review.

\begin{itemize}
    \item \textbf{RQ1:} What types of SSD do AI-based automated speech therapy tools address?
    \item\textbf{RQ2:} What is the level of autonomy achieved by such tools?
    \item \textbf{RQ3:} What are the different modes of intervention (delivery modes: mobile,
computer, robots, etc., and presentation modes: games, storytelling, etc.)?
    \item \textbf{RQ4:} How effective are such tools with respect to the conventional mode of
speech therapy provided by SLPs?
\end{itemize}

\subsection{Eligibility Criteria}
\subsubsection{Types of Studies}
We considered full articles, review papers, and short papers which proposed automated
speech therapy tools using AI techniques such as machine learning and deep learning.
The studies were restricted to articles written in English and published from 2007 to
2022 (research carried out over the last 15 years). The objective is to provide a snapshot
of the research domain. We carried out the final search on February 4th, 2022.
\subsubsection{Types of Participants}
We included participants of any age and gender with SSD, such as articulation disorder, phonological disorder, apraxia, dysarthria, cleft palate, and hearing impairment.
However, we excluded studies specifically addressing cognitive conditions such as intellectual disability, Alzheimer’s disease, Down’s syndrome, Parkinson’s disease, and
Autism Spectrum Disorders.
\subsubsection{Types of Interventions}
We didn’t restrict on types of intervention and included all studies that proposed automated speech therapy using different intervention methods. Studies included robotics-based, mobile-based, computer-based interventions along with gamified and storybased intervention methods.

\subsection{Information Sources}
The studies were identified by searching electronic databases using the search term
generated using keywords from the research questions. The search string was applied
to Scopus, IEEE Xplore, and ACM Digital Library electronic databases. The results
from the databases were extracted, and inclusion/exclusion criteria were applied to
find relevant studies

\subsection{Search Terms}
The following keywords were used to search all the databases: speech, language, disorder, impairment, assessment, therapy, rehabilitation, treatment, AI, artificial intelligence, automated, automatic. Boolean operators were used to combine the terms as:
(”AI” OR ”Artificial Intelligence” OR ”automa*”) AND (”speech” OR ”language”) AND
(”disorder” OR ”impairment”) AND (”assessment” OR ”therapy” OR ”rehabilitation”
OR ”treatment”).

\subsection{Study Selection and Data Collection}
\begin{figure}[htp]
    \centering
    \includegraphics[width=\textwidth]{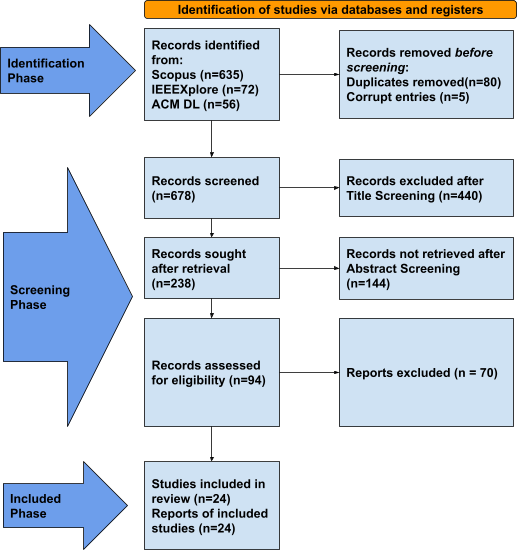}
    \caption{Prisma Systematic Review Process applied to 763 papers}
    \label{fig:prisma_flowchart}
\end{figure}

We found a total of 763 research studies from individual databases, i.e., 635 from
Scopus, 72 from IEEE Xplore, and 56 from the ACM Digital Library. Then we removed
duplicates and corrupt entries to find 678 papers for the screening phase. We performed
the screening of the studies in three stages. At the first stage, two authors screened
the titles, which resulted in 238 research studies. Three authors reviewed the abstracts
in the next screening stage, which resulted in 94 research studies. Finally, all five
authors reviewed the full texts of the 94 articles. After applying the inclusion and
exclusion criteria for this systematic literature review, we selected 24 research studies.
Disagreements during the screening process were resolved by discussion and voting by
all the authors.
Figure \ref{fig:prisma_flowchart} shows all the phases: ”identification,” ”screening,” and ”included” according
to the PRISMA protocol \citep{10_shamseer2015preferred}.

\section{Results}
We report our results in five different sections. In the first section 3.1, we analyzed
the included papers with respect to paper counts, authors, regions, languages, and
venues. In the following sections (see 3.2,3.3,3.4,3.5), we present our findings based on
the research questions addressed in this systematic literature review.

\subsection{Paper Counts, Authors, Regions, Languages, and Venues}
\begin{figure}[htp]
    \centering
    \includegraphics[width=\textwidth]{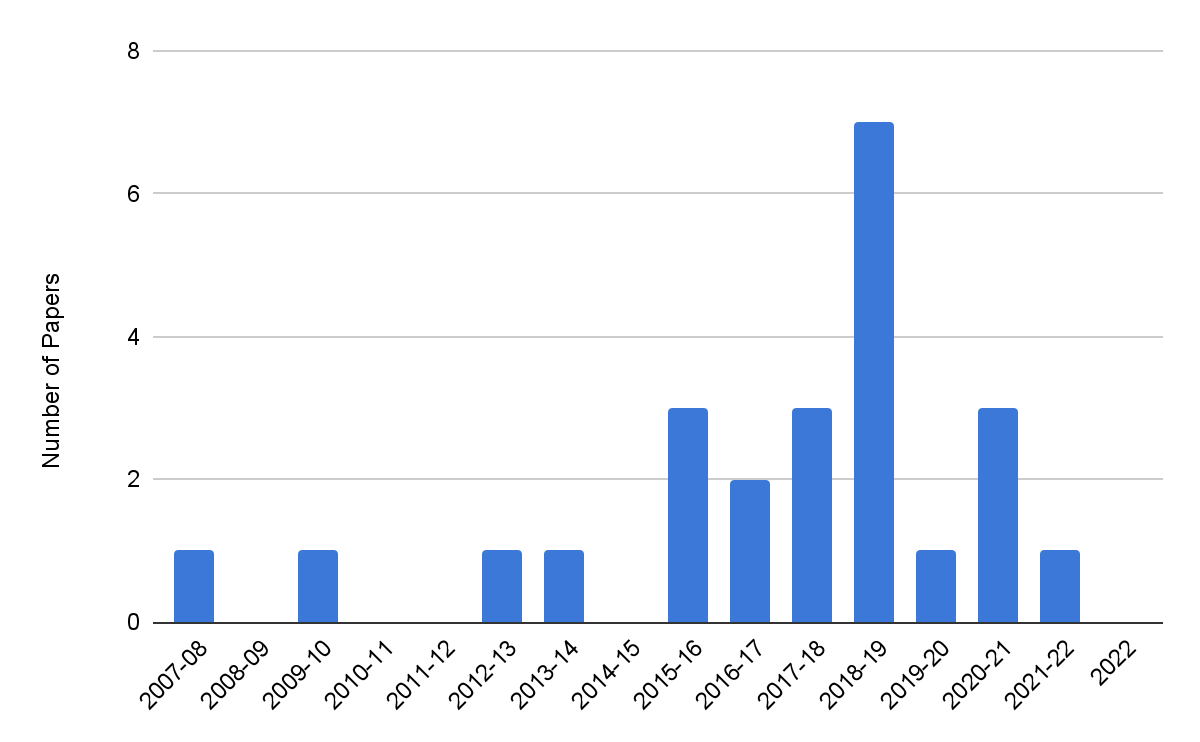}
    \caption{Number of papers according to the year of publication}
    \label{fig:publication_yearwise}
\end{figure}

The final papers selected after the full review were 24 papers from 23 different venues.
The number of studies on AI-Based automated speech therapy tools for persons with
SSD published shows an upward trend over the years (see Figure \ref{fig:publication_yearwise}). Out of 24 papers,
we can observe that 20 articles were published during the last 6-7 years.

\begin{figure}[htp]
    \centering
    \includegraphics[width=\textwidth]{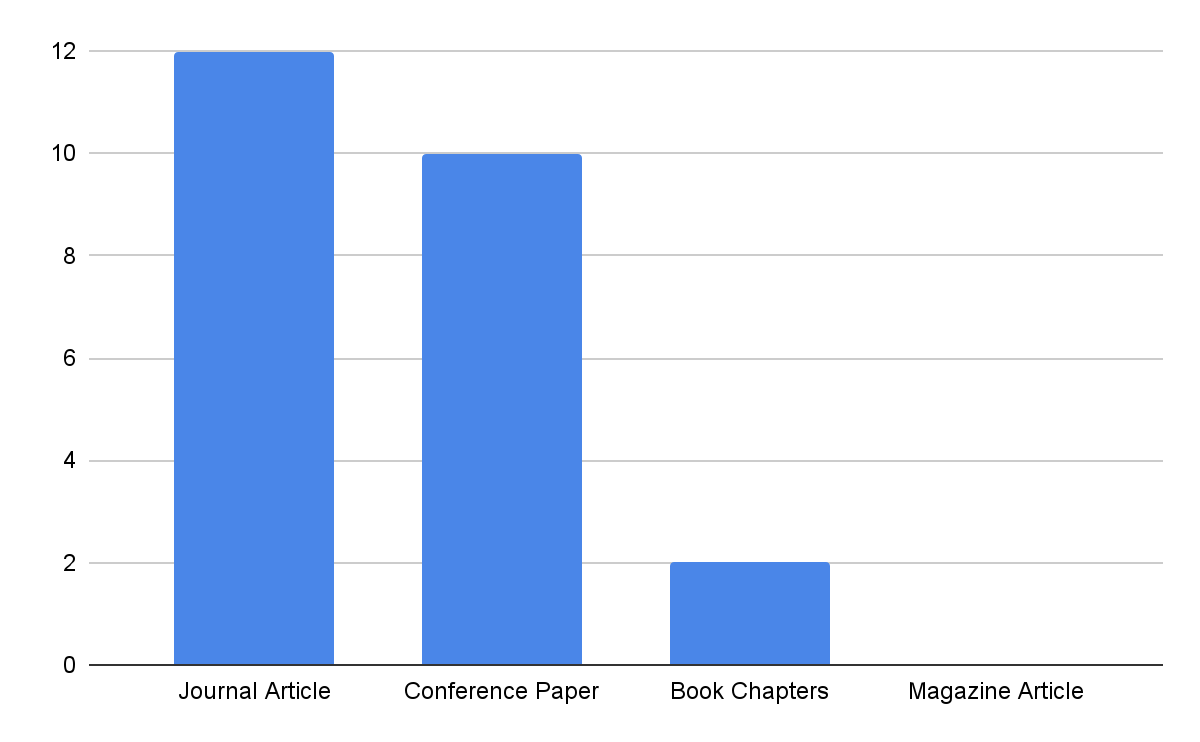}
    \caption{Number of publication type}
    \label{fig:publication_type}
\end{figure}

The majority of the paper included in this study were published in journals (see
Figure \ref{fig:publication_type}). Additionally, there were ten papers published in conference proceedings and
two book chapters among the 24 included studies. However, we could not find any
eligible studies published in a magazine.

\begin{figure}[htp]
    \centering
    \includegraphics[width=\textwidth]{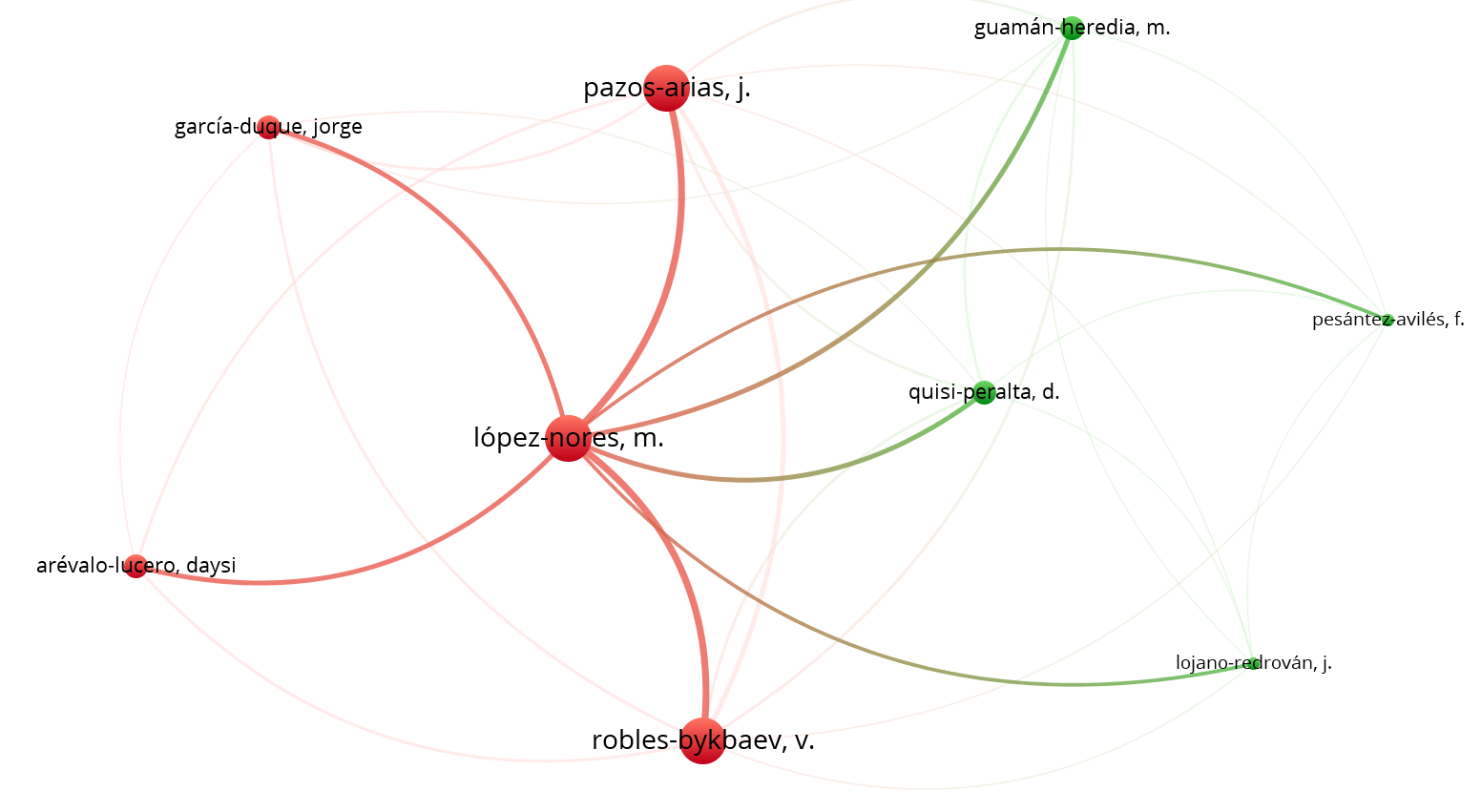}
    \caption{Main cluster of co-authorship analysis}
    \label{fig:co_authorship_analysis}
\end{figure}

\begin{figure}[htp]
    \centering
    \includegraphics[width=\textwidth]{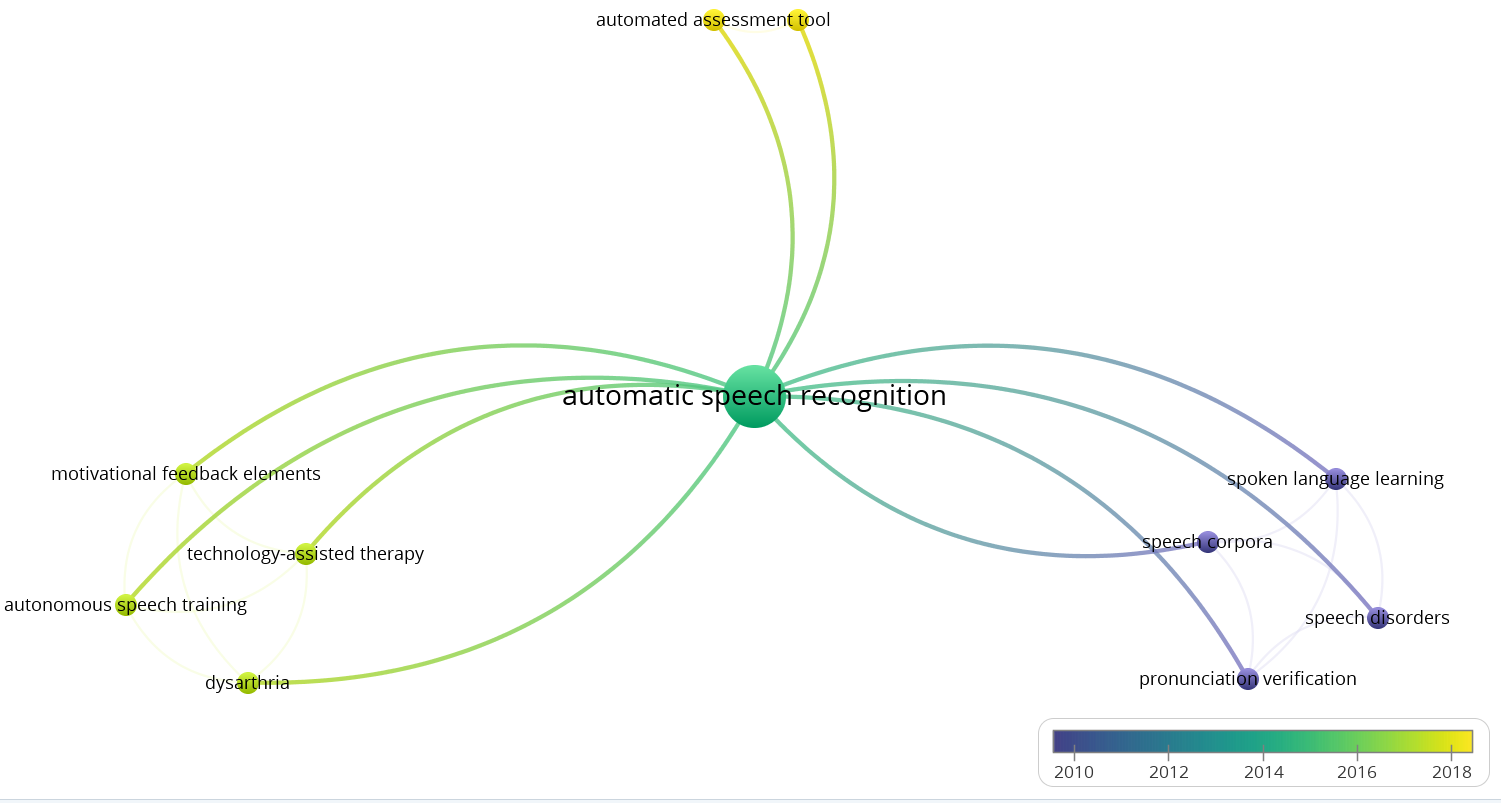}
    \caption{Main cluster of co-occurrences of keyword analysis}
    \label{fig:cluster_keyword}
\end{figure}

There were 91 unique authors identified from the included studies. The VOSviewer software was used to calculate the most impactful authors, generate co-authorship clusters, and perform co-occurrences of keyword analysis \citep{11_vos}. All the authors were counted irrespective of the authorship order, with the same weightage applied to all the authors. However, high weightage was attributed to authors publishing more articles. In addition, to find the list of most impactful authors, their collaborating links were also considered, along with the number of published documents. The top ten most impactful authors are listed in the Table 1 . The most significant cluster of authors based on the number of articles and collaborative link strength was found, as shown in the Figure \ref{fig:co_authorship_analysis}. It is worth noting that 79 authors (86.81 \%) contributed to only one paper in the included studies, i.e., have only one work relating to AI-based automated speech therapy in the last 15 years. Moreover, after analyzing the author’s keywords of the included studies, the most significant cluster of linked and co-occurred keywords was found as shown in the Figure \ref{fig:cluster_keyword}. The most significant keyword was ASR(Automatic Speech Recognition).

% Add the most impactful author Table

\begin{figure}[htp]
    \centering
    \includegraphics[width=\textwidth]{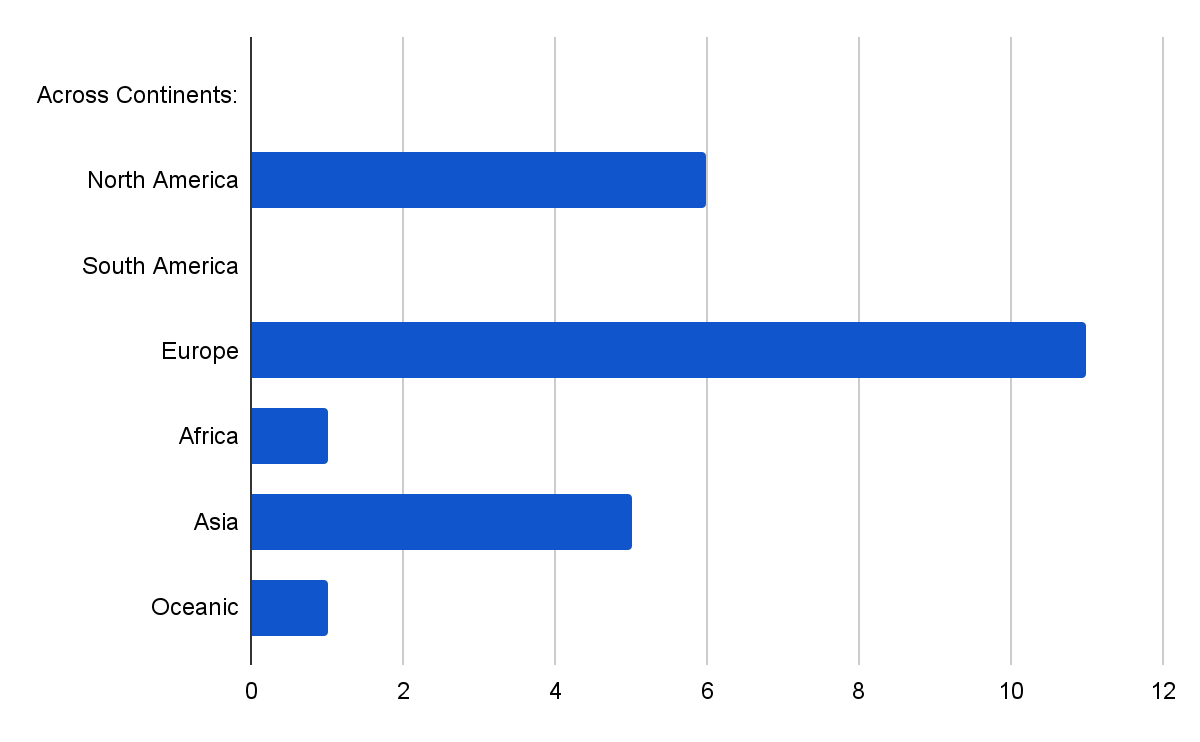}
    \caption{Geographical distribution of the papers}
    \label{fig:geographical_distribution}
\end{figure}

We further report the geographical distribution of the included studies based on the
location of the study indicated in the paper (see Figure \ref{fig:geographical_distribution}). We looked at the author’s affiliation and funding agency when required. Most papers reported on studies which
were conducted in Europe (11 papers) and North America (6 papers). Studies conducted in Europe include four studies from Spain and one study each from Germany, Hungary, Romania, Portugal, the Czech Republic, and Italy. On the other hand, studies from North America include four studies from the USA, one collaborative study between Panama and Nicaragua, and another study from Mexico. Moreover, five papers reported studies in Asia, which includes China (2 studies), India (1 study), Taiwan (1 study), and the Philippines (1 study). However, other continents are heavily underrepresented; Africa and Oceanic each have one study conducted. Finally, we could not find any eligible studies meeting our selection criteria which were conducted in South America.

\begin{figure}[htp]
    \centering
    \includegraphics[width=\textwidth]{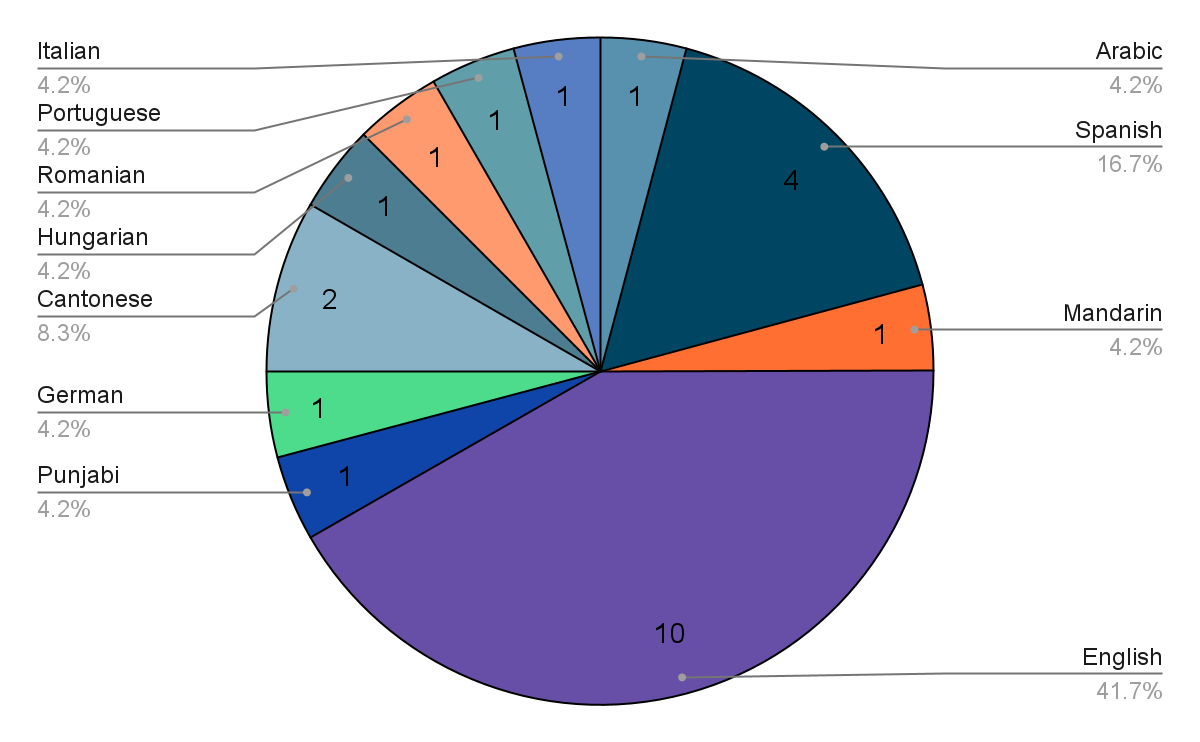}
    \caption{Language distribution of the papers}
    \label{fig:language_distribution}
\end{figure}

We presented the language distribution of the papers based on the language addressed by the AI-based automated speech therapy tools as reported in the studies (see Figure \ref{fig:language_distribution}). The most addressed languages were English (10 studies) and Spanish (4 studies). Furthermore, two studies addressed the Cantonese language, and there was only one study each for Punjabi, German, Hungarian, Romanian, Portuguese, Italian, Arabic, and Mandarin. The studies were drawn from 23 unique venues. We could observe that the vast majority of venues from which papers were chosen (95.65\%) were represented by only one article. Only one venue, i.e., ”Studies in Health Technology and Informatics,” had published two papers included in this review.

\subsection{Speech Sound Disorders (RQ1)}
\begin{figure}[htp]
    \centering
    \includegraphics[width=\textwidth]{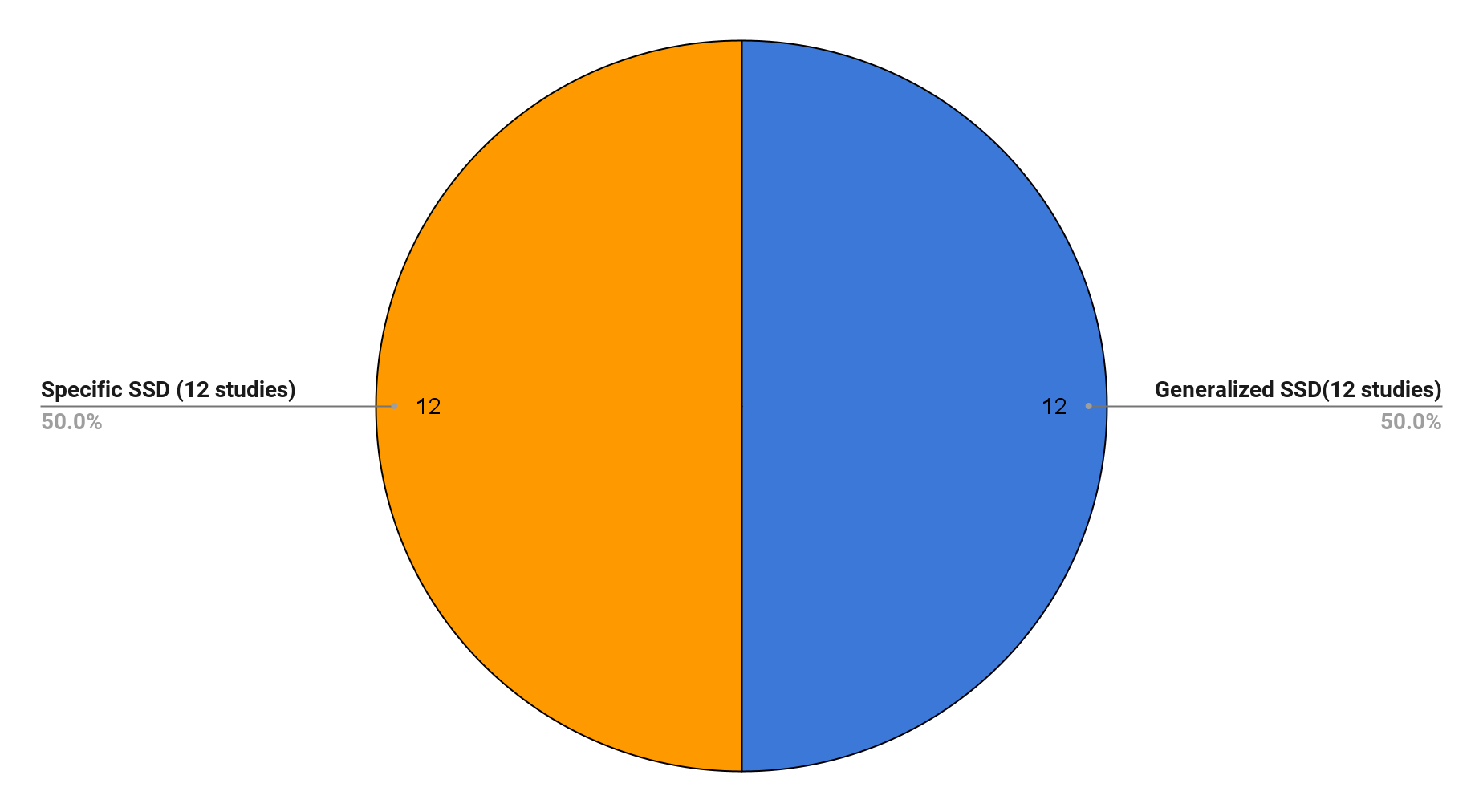}
    \caption{Distribution of papers addressing generalized and specific SSD}
    \label{fig:generalized_specific}
\end{figure}

We found that researchers have addressed multiple types of SSD in the literature. However, 12 studies out of 24 studies did not address any specific SSD (see Figure \ref{fig:generalized_specific}). These studies proposed automated tools for a generalized SSD population and experimented without specifying any particular SSD \citep{8_robles2017onto,12_desolda2021system,13_ng2020cuchild,14_ng2018automated,15_mahmut2018computer,16_das2017automated,17_robles2016ontology,18_robles2016evaluation,19_robles2015spelta,20_seddik2013computer,21_duval2020approaches,22_samonte2018assistive}. Researchers have also specifically worked and devised AI-based tools for persons with hearing impairment \citep{23_sztaho2018computer,24_cespedes2015sega}.A novel tongue-based Human Computer Interaction tool \citep{25_bilkova2020human} and gamified AI-based tool \citep{7_duval2018spokeit} for persons with motor speech disorder have been proposed. 

\begin{figure}[htp]
    \centering
    \includegraphics[width=\textwidth]{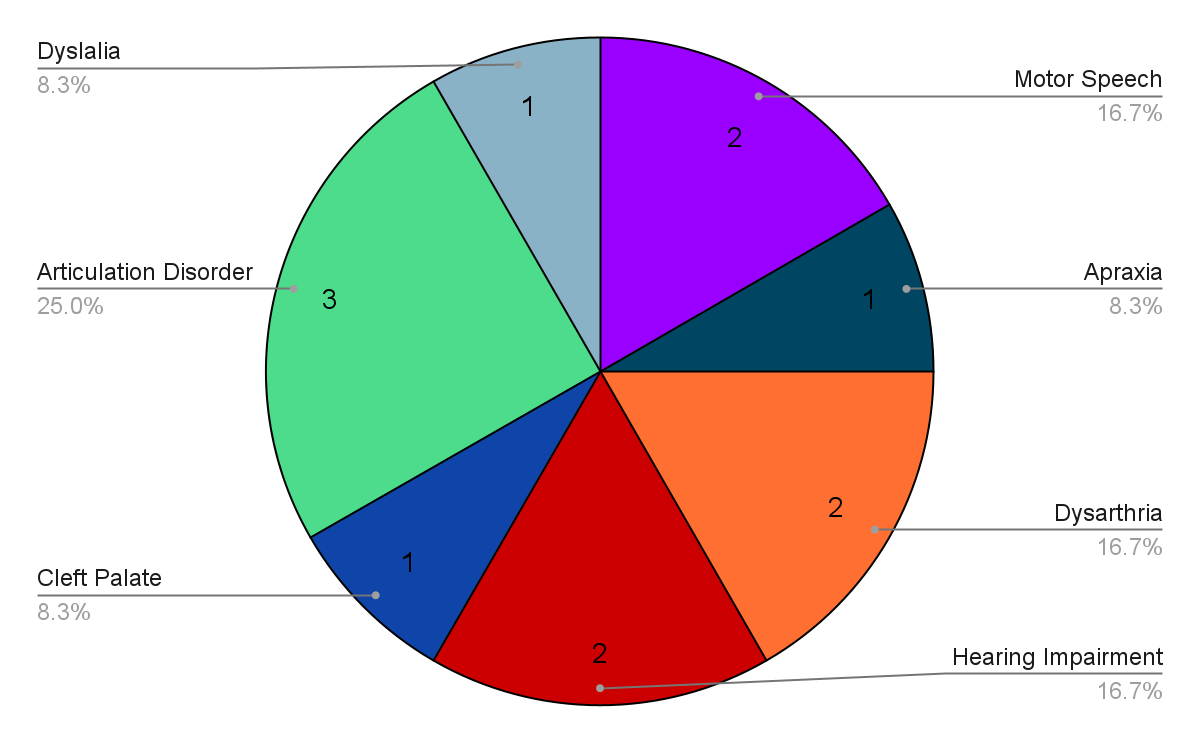}
    \caption{Distribution of papers addressing specific SSD}
    \label{fig:distribution_specific_SSD}
\end{figure}

Moreover, Frieg et al. proposed a digital training system for dysarthric patients \citep{26_frieg2017isi}. In another similar study, Saz et al. devised ASR-based tools and technologies and conducted user studies specifically for dysarthric patients \citep{27_saz2009tools}. Singh et al. and Chen et al. developed and assessed automatic AI-based speech therapy tools for articulation disorder in Punjabi and Mandarin, respectively \citep{28_singh2015automatic, 29_chen2007development}. Ballard et al. conducted a feasibility study of a tablet-based automated feedback tool for apraxia patients \citep{30_ballard2019feasibility}. On the other hand, Ramamurthy et al. developed a novel companion robot, ”Buddy,” for cleft lip and palate disorder children \citep{31_ramamurthy2018buddy}. In another study, Rivas et al. proposed using a virtual world to provide speech therapy for children with dyslalia \citep{32_rivas2012proposal}. It is worth noting that only one study was related to speech data collection for Cantonese to perform phonology and articulation assessment \citep{13_ng2020cuchild}. The Figure \ref{fig:distribution_specific_SSD} shows the distribution of papers addressing specific SSD.

\subsection{Level of Autonomy (RQ2)}
Researchers worldwide have amplified the debate between autonomy vs. human control due to the risks and concerns associated with AI and large-scale automation \citep{33_shneiderman2020human}. In this area concerning automation and AI in speech therapy, we studied the level of autonomy achieved by AI-based automated speech therapy tools. In many studies, researchers build fully automated AI-based speech therapy tools without considering the role of parents, SLPs, and other stakeholders. While Desolda et al. emphasized the role of caregivers and SLP in the design of a remote therapy tool, ”Pronuntia” \citep{12_desolda2021system}, Ng et al. proposed a fully automated assessment tool using the CUChild 127 speech corpus in Cantonese \citep{14_ng2018automated}. In another study, Bilkova et al. developed a novel lip, tongue, and teeth detection system using Convolutional Neural Network (CNN) and Augmented Reality (AR) for supporting the automatic evaluation of speech therapy exercises \citep{25_bilkova2020human}.
\\Furthermore, Sztaho et al. proposed a fully automated speech therapy tool by displaying visual feedback on intensity(accent), intonation, and rhythm to children with hearing impairments \citep{23_sztaho2018computer}. In another similar study, Hernandez et al. developed a serious game with an automatic feedback feature for hearing impaired children \citep{24_cespedes2015sega}. Ballard et al. performed a feasibility study of their tablet-based, fully automated therapy tool for children with apraxia without any role of SLP and other stakeholders \citep{30_ballard2019feasibility}. Moreover, V. Robles-Bykbaev et al. proposed a framework imitating the main functionality of SLP along with a robotic assistant motivating children in therapy activity and automatically giving real time feedback \citep{8_robles2017onto,17_robles2016ontology,18_robles2016evaluation,19_robles2015spelta}. In another similar study, Ramamurthy et al. proposed a companion robot, ”Buddy,” which automatically evaluates speech exercises of children with CL/P disorders with the feature of monitoring by SLPs \citep{31_ramamurthy2018buddy}.

\subsection{Modes of Intervention (RQ3)}
\begin{figure}[htp]
    \centering
    \includegraphics[width=\textwidth]{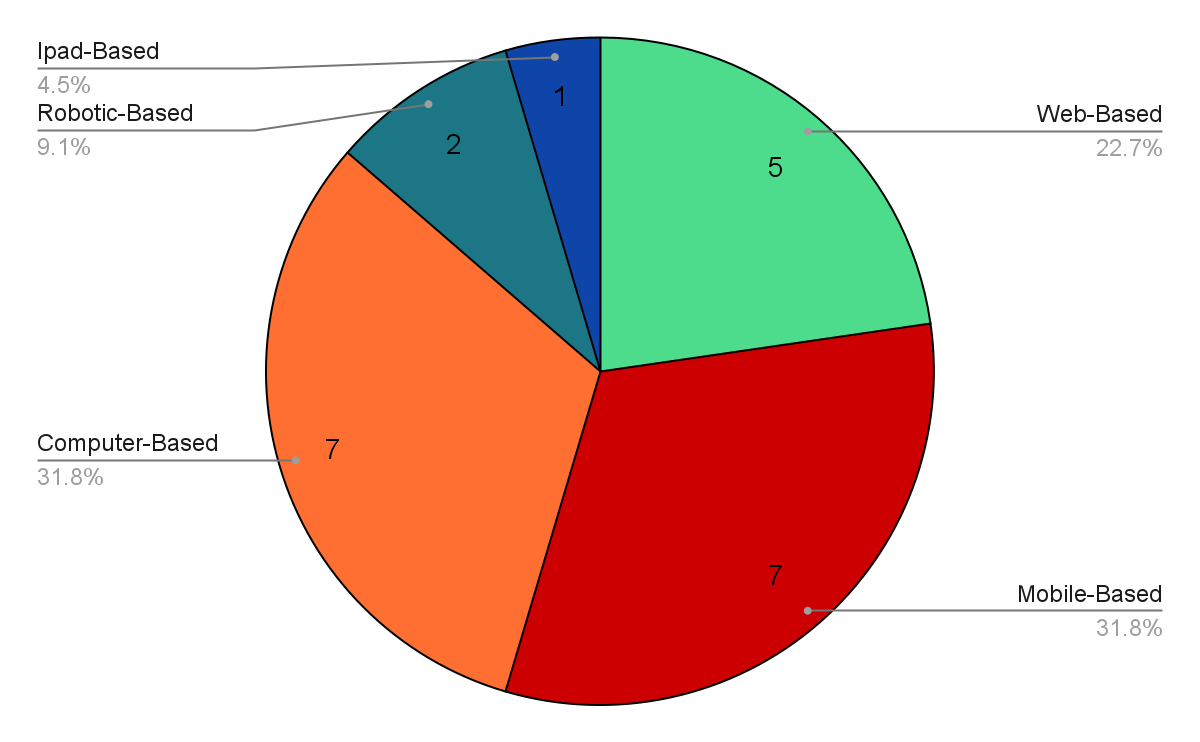}
    \caption{Distribution of papers based on modes of intervention}
    \label{fig:distribution_modes_intervention}
\end{figure}

Researchers have adopted different modes of intervention while implementing AI-based automated speech therapy tools for persons with SSD (see Figure \ref{fig:distribution_modes_intervention}). As these therapies are often targeted at children, researchers emphasize developing tools that trigger excitement and build companionship. Desolda et al. proposed a web application for children, SLPs, and caregivers, allowing SLP to assign therapy exercises to children with SSD \citep{12_desolda2021system}. The system automatically evaluates the correctness of the exercises and gives real time feedback. On the other hand, Ballard et al. proposed a tabletbased therapy tool for children with apraxia \citep{30_ballard2019feasibility}. Furthermore, Ng et al. and Sztaho et al. proposed a computer-based prosody teaching system for children with hearing impairment and a computer-based visual feedback system for the hearing impaired, respectively \citep{14_ng2018automated,23_sztaho2018computer}. Bykbaev et al. proposed a novel robotic assistant along with a fully automatic framework imitating the work of SLP \citep{8_robles2017onto}. In another similar study, Ramamurthy et al. proposed a therapy robot, ”Buddy,” allowing children to practice assigned exercises at home \citep{31_ramamurthy2018buddy}. Many studies have incorporated serious games as an intervention tool for automatic speech therapy \citep{7_duval2018spokeit,21_duval2020approaches,25_bilkova2020human,31_ramamurthy2018buddy,34_anjos2017serious}. One of the studies incorporated augmented reality to build a serious game using tongue detection \citep{25_bilkova2020human}.

\subsection{Effectiveness (RQ4)}
\begin{figure}[htp]
    \centering
    \includegraphics[width=\textwidth]{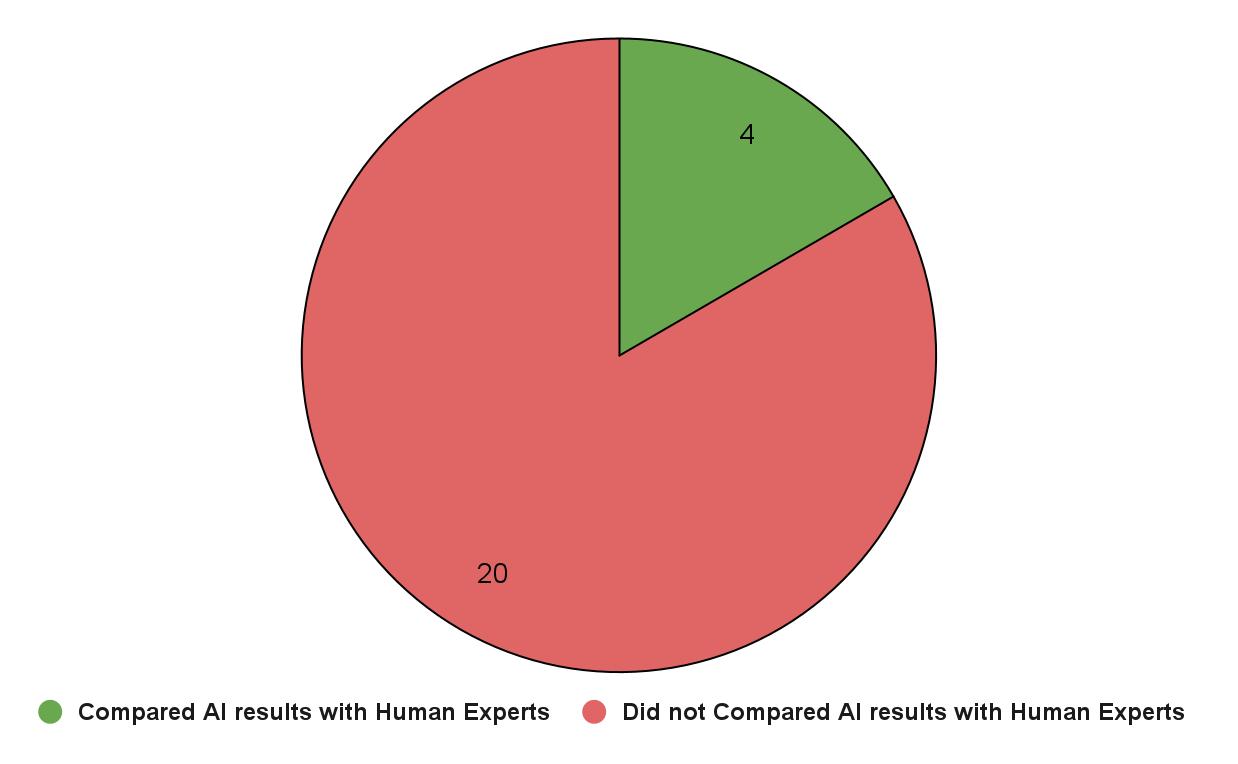}
    \caption{Distribution of papers comparing AI results with human experts}
    \label{fig:distribution_AI_human_experts}
\end{figure}

The effectiveness of AI-based automated speech therapy tools depends on their performance compared to the conventional mode of speech therapy provided by SLPs. Moreover, automated speech therapy tools providing wrong feedback can be disastrous to children’s speech improvement. Few studies (4 out of 24) compared the results of their automated tool with the conventional mode of speech therapy provided by SLPs (see Figure \ref{fig:distribution_AI_human_experts}). Ballard et al. conducted an interrater agreement test between their ASR tool and SLPs and found ASR-human agreement averaged 80\% \citep{30_ballard2019feasibility}. In another study, Sztaho et al. found that their automated tool scores correspond to the subjective evaluation by SLPs \citep{23_sztaho2018computer}. Bykbaev et al. found that over 90\% of the therapy plans generated automatically by their expert ”Spelta” were ”better than” or ”as good as ” what the SLPs would have created manually \citep{18_robles2016evaluation}. Moreover, in the study by Saz et al., their Automatic Speech Recognition (ASR) and Pronunciation Verification (PV) modules based on impaired speech utterances provided performance similar to SLPs \citep{27_saz2009tools}.

\section{Discussion}
We conducted this systematic literature review based on a sample of 24 out of 678 research papers deriving from Scopus, IEEEXplore, and ACM DL databases. Exciting insights and trends emerged from our analysis of these papers. In recent years, we observed an increasing interest in AI-based auto- mated speech therapy tools. This growing interest can be due to the recent advancement in ASR technology and its improved accuracy. Surprisingly, 79 authors (86.81\%) out of 91 unique authors have only one work on AI-based automated speech therapy in the last 15 years. This data suggests that a significant amount of research in this field is ”one-off” by authors. Most authors explored the research area with one idea and did not develop or evaluate it further. We found that ”automatic speech recognition” is the most emphasized keyword by the authors. This finding is consistent with the notion that ASR is the core of AI-based automated speech therapy tools. The majority of studies were from European, North American, and Asian countries, and the most prevalent language targeted by the included studies was English. This finding is in line with the fact that English is the most widely adopted language for ASR technologies \citep{35_benzeghiba2007automatic}. However, we can also observe that researchers have attempted to build AI-based automated speech therapy tools in other languages.

Furthermore, we found that articulation disorder was the most frequent disorder addressed by the included studies, with three studies dedicated to them. This may be due to the fact that articulation disorder are commonly found in persons with other SSD \citep{2_flipsen2015}. The results show that most studies aimed at developing fully automated speech therapy tools without considering the role of other stakeholders such as speechlanguage pathologists, caretakers, parents, and family members. This finding is in line with the widespread belief that researchers traditionally follow the one-dimensional framework of levels of automation by Sheridon and Verplank, which suggests that more automation leads to less human control and vice versa \citep{36_sheridan1978human}. Moreover, a fully automated system may bring multiple concerns, such as biased data, privacy, replacement of jobs, and extreme automation may lead to disastrous consequences \citep{33_shneiderman2020human}. In the case of AI-based automated speech therapy, many concerns arise, including biased speech data, replacement of SLPs, and privacy of children’s speech data. We further found that mobile-based deployment of AI-based automated speech therapy was more common among the included studies. A possible explanation is that researchers are more interested in building affordable and accessible automated speech therapy tools. Another significant issue we observed is that few studies compared their automated tools’ results with human experts such as SLPs. This considerable insight questions the effectiveness of automated AI-based speech therapy tools compared to expert SLPs. 
\\There are some limitations in our study which is worth mentioning. We relied on three databases: ACM DL, IEEE Xplore, and Scopus; therefore, we may have missed relevant papers published in other databases. Another limitation is the inapplicability of quality appraisal methods such as the ”Risk of Bias Assessment” in our study, as in the case of health sciences. Furthermore, our study restricts papers addressing SSD defined by ASHA and excluded studies addressing other related disorders.

\section{Conclusion}
This systematic literature review was based on the PRISMA Statement to analyze papers on AI-based automated speech therapy tools for persons with SSD. We extracted relevant data from the included articles based on four predefined research questions: Types of SSD addressed; Level of autonomy achieved by such tools; Modes of interventions and Effectiveness of such tools. Our study answers all the predefined research questions providing a snapshot of the research carried out in the domain. We found that articulation disorder, hearing impairment, dysarthria, and motor speech were the most frequently studied disorders, addressed in three, two, two, and two studies (research question 1). However, 50\% of the studies did not address any specific SSD. Concerning the level of autonomy (research question 2), almost all studies proposed fully automated AI-based speech therapy tools suggesting that researchers did not emphasize the role of caretaker, parents, family members, and SLPs. Addressing the modes of intervention (research question 3), most researchers proposed mobile-based and computer-based applications. Finally, the analysis of the effectiveness (research question 4) of such AI-based speech therapy tools provides us the insights that very few studies have compared their proposed system’s effectiveness with expert SLPs.
\\Based on the findings and insights from our research questions, we propose the following directions for future research on AI-based automated speech therapy tools.
\begin{itemize}
    \item Development of speech corpora and AI-based automated speech therapy tools for
under-represented languages and its deployment in under-developed regions where the
shortage of SLPs prevails.
    \item Development of such tools for specific SSD instead of generalized SSD.
    \item Implementation of a Human-Centered AI approach in developing such tools, i.e., involving all stakeholders in the design process instead of focusing on developing fully automated tools.
    \item Conducting usability studies for understanding the effectiveness of different modes of delivery (mobile device, computer, companion robots) and different modes of presentation (gamified content, storytelling).
    \item Development of a robust framework for measuring the effectiveness of such tools with respect to conventional speech therapy.
\end{itemize}

\section{Conflict of Interest}
On behalf of all authors, the corresponding author states that there is not conflict of interest.

\bibliographystyle{apacite}
\bibliography{interactapasample}

\end{document}